# Optimization of light trapping micro-hole structure for high-speed high-efficiency silicon photodiodes

Ekaterina Ponizovskaya Devine,[1,3] Hilal Cansizoglu,[1] Yang Gao,[1] Kazim G. Polat,[1] Soroush Ghandiparsi,[1] Ahmet Kaya,[1] Hasina H. Mamtaz,[1] Ahmed S. Mayet,[1] Yinan Wang,[1] Xinzhi Zhang,[1] Toshishige Yamada[2,3], Aly F. Elrefaie,[1,3] Shih-Yuan Wang,[3] M. Saif Islam[1]

**Affiliations:**

[1]Electrical and Computer Engineering, University of California, Davis, Davis, CA 95618.

[2]Electrical Engineering, Baskin School of Engineering, University of California, Santa Cruz, Santa Cruz, CA 95064.

[3]W&WSens Devices, Inc., 4546 El Camino, Suite 215, Los Altos, CA United States.

**Abstract**

We optimized micro-holes in a thin slab for fast Si photodetectors at wavelength 800-950nm. Lateral modes are shown to be responsible for the effective light trapping. Small disorder and cone hole shapes helped achieve uniform quantum efficiency in a wide wavelength range.

**Summery**

The Si *pin* PDs at the short-reach multimode datacom wavelengths of 840-860 nm and at a short wavelength division multiplexing (SWDM) band of 850-950 nm could facilitate the integration [1] and reduce the cost. Our recent study [2] showed experimentally that a microstructure can increase the Si PDs quantum efficiency (QE) so that less than 2 micron intrinsic (i) layer can show the QE>50% with data rate 20 Gb/s for wavelength 800-950nm . In this study we analyze the modes in the structure numerically in order to optimize shape and size of the holes for better QE.

This study presents Finite Difference Time Domain (FDTD) simulation shows that the lateral modes that propagates along the PDs surface are responsible for the light trapping. Optimizing the structure we found that the holes comparable to the wavelength provide the better bending of the light into the lateral modes. The designed PDs have 2D periodic micro-holes, whose view from the top and cross-sections are in the *x-z* plane are shown in Fig. 1a. The structure consist of p, i and n layers of Si on a SOI substrate. The hole array supports a set of modes with wave vectors in *z* direction, $k_z$ and $k_c$ in *x-y* plane. Each hole couples the vertically incident light into the x-y plane. The lateral modes with larger $k_c$ can have full reflection from the bottom of the PD and they can form a modes guided in the x-y plane. The guided resonance modes [3] in xy plane causing a bigger absorption. The FDTD simulation for the holes with period 500nm and diameter of 400nm shows small absorption with spikes at the wavelengths of guided resonance (Fig.1b, blue line). However, the leaky modes can be useful for the absorption if $Im(k_l)<\square$, where $k_l$ is the wave vector of the leaky mode and $\square$ is the Si bulk absorption. At this condition the mode can be successfully absorbed before it will leak outside the structure. The longer the light propagate into *i*-region and the better it confined in it the better is the absorption for a wider range of the wavelength. As we can see from FDTD simulations the bigger hole with 800nm period and 700nm diameter (Fig1b, green line) had bigger absorption and the tapered holes, that have cone shapes with period 1000nm and 800nm, produce the highest absorption(Fig.1, red line). The cone shapes had smaller reflection due to the effect is similar to the Lambertian reflector that helps to trap light in Si [4]. They also provide better absorption with smaller resonant features in the 800-900 nm range.

The Poynting vector from a single hole and the hole array is presented on Fig.2a,b. for the wavelength 850nm. The Poynting vector shows that the hole produces the lateral waves in x-y around the nanoholes.

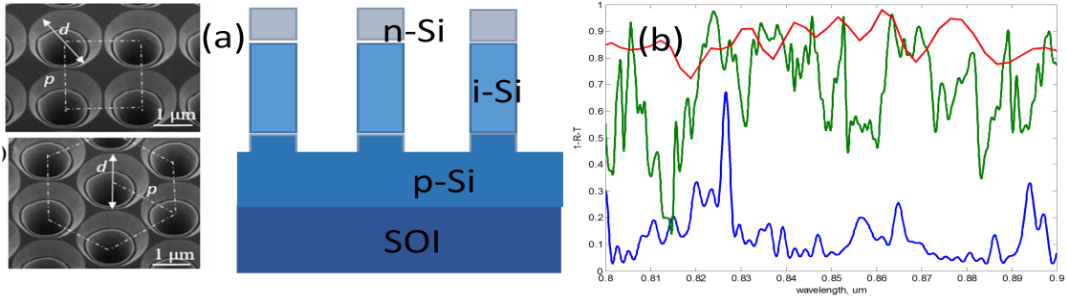

Fig.1 (a) view of the micro-hole structure from the top, in the xz plane; (b) FDTD simulation for absorption for cylindrical holes with period 500nm (blue), 800nm(green) and tapered holes with period 1000nm.

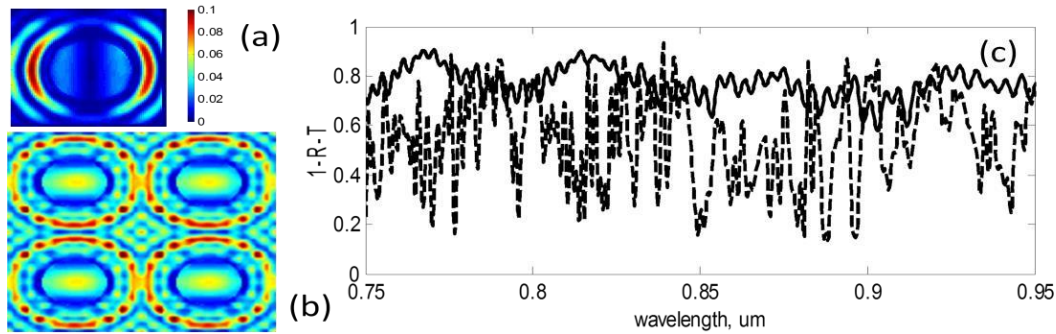

Fig.2 (a) The FDTD simulations shows the Poyniing vector in lateral (xy) plane from a single hole, (b) Poynting vector for a hole array; (c) absorption for periodical cylindrical holes (dashed line) and for same hole array with disorder (solid line).

The disorder in the periodical structure destroys the guided resonances and produces more lateral modes that propagates long enough in *x-y* plane to be absorbed in I layer. The Fig2c shows the absorption for disordered array of cylindrical holes (solid line) and periodic array of the same cylindrical holes. Periodic array have less absorption and more resonance behavior. In real structure the absorption depends also on the thickness of p and n layer. Only the part of the light that absorbs in i-layer contributes in QE. QE was estimated based on the FDTD more than 60% for 2 micron i-layer, the results are in agreement with the experimental data.

Our research showed that micro/nano structure and the effect of lateral modes propagation can improve QE and can help to design more efficient photodetectors in the longer wavelength using Ge or GeSi alloys as well as avalanche photodiods.